\RequirePackage{ifpdf}
\ifpdf
\else
\PassOptionsToPackage{dvipdfmx}{graphicx}
\PassOptionsToPackage{dvipdfmx}{hyperref}
\fi

\documentclass[12pt,a4paper]{article}
\usepackage{jheppub}
\allowdisplaybreaks[4]
\usepackage{amsmath,bm}
\usepackage{subcaption}
\usepackage{slashed}
\usepackage{mathrsfs}

\usepackage{color}

\newcommand{\pa}{\partial}

\newcommand{\pmat}[1]{\begin{pmatrix} #1 \end{pmatrix}}

\title{
Singularity avoidance in black hole interiors by quantum gravity effects
}

 \author[a]{Takeshi Chiba,}
 \author[a,b]{Hiroki Matsui,}
 \author[a]{and Keiju Murata}
  \affiliation[a]{Department of Physics, College of Humanities and Sciences, Nihon University, Sakurajosui,
 Tokyo 156-8550, Japan}
 \affiliation[b]{Center for Gravitational Physics and Quantum Information, 
Yukawa Institute for Theoretical Physics, Kyoto University, 606-8502, Kyoto, Japan}
\emailAdd{matsui.hiroki@nihon-u.ac.jp}
\emailAdd{murata.keiju@nihon-u.ac.jp}

\abstract{
The quantum nature of the Schwarzschild black hole interior is investigated through the Wheeler-DeWitt (WDW) equation.
The interior of a static, spherically symmetric black hole is described by the Kantowski-Sachs (KS) metric, which represents a homogeneous but anisotropic cosmology. We derive the Hamiltonian for the gravitational system corresponding to the black hole interior and obtain the associated WDW equation. By varying the gravitational constant as a parameter controlling quantum effects, we examine how the solutions of the WDW equation change with respect to this parameter. In the parameter regime where quantum effects are negligible, we find that the wave packet solutions closely follow the classical trajectory of the black hole interior. 
On the other hand, as quantum effects are enhanced, 
the wave packet deviates from the classical trajectory and exhibits behavior suggestive of singularity avoidance. To quantify this behavior, we introduce an appropriate "clock" inside the black hole and compute the time to singularity formation with respect to this clock. The results show that stronger quantum effects lead to a longer formation time, suggesting a tendency toward the avoidance of singularity formation due to quantum gravity effects. 
}

\begin{document}
\maketitle

\section{Introduction}
The unification of general relativity with quantum mechanics, i.e., the construction of a consistent quantum theory of gravity, remains one of the central open problems in theoretical physics. A significant challenge is the presence of spacetime singularities, where the classical framework of general relativity unavoidably breaks down. Cosmological spacetimes, which contain initial singularities, and black holes, whose interiors contain singularities, thus serve as natural laboratories for exploring the quantum structure of spacetime and testing concepts in quantum gravity.

A central role in investigating quantum gravitational effects in black holes and cosmology is played by the Wheeler-DeWitt (WDW) equation~\cite{PhysRev.160.1113}. Its solution is given as the wave function of the universe, expressed as a functional of the metric. However, the configuration space is infinite-dimensional, making its analysis extremely challenging. A common approach to this difficulty is the use of the minisuperspace model, in which symmetries are imposed on space-time so that the dimensionality of the configuration space is reduced to a finite one. Although the minisuperspace model is not a systematic "approximation" in the strict sense, it is widely studied as a toy model that retains certain aspects of the full theory. In quantum cosmology, the resolution of the initial singularity has long been discussed on the basis of the WDW equation within the minisuperspace model. (See, for example, \cite{Halliwell:1989myn} for a comprehensive review of quantum cosmology and the references therein.)

Techniques developed in quantum cosmology have also been applied to the study of the interior of quantum black holes.
Inside a static black hole, the time and spatial coordinates exchange their roles compared to the exterior. In other words, the black hole interior becomes a dynamical spacetime. 
The best-known example is the Schwarzschild spacetime, whose interior region is isometric to the Kantowski-Sachs (KS)  spacetime: 
a homogeneous and anisotropic spacetime with a spatial topology of ${\mathbb R}\times S^2\times S^1$ ~\cite{Kantowski:1966te}. 
Thus, for static black holes, canonical quantization can be studied within the minisuperspace model in the same manner as in cosmology.
Indeed, many studies have addressed the black-hole singularity problem by applying the WDW equation to the KS spacetime~\cite{Nambu:1987dh,Nakamura:1993nq,Bouhmadi-Lopez:2019kkt,Perry:2021mch,Brahma:2021xjy,Perry:2021udd,Hartnoll:2022snh,Kan:2022ism,Blacker:2023oan,Piazza:2025uxm}. 
In these pioneering studies, attention has been focused primarily on semiclassical solutions inside black holes, and solutions of the WDW equation in the fully quantum regime have not been systematically explored. In the present work, we systematically analyze quantum gravitational effects by varying the gravitational constant, which serves as a parameter controlling quantum effects, and study the wave function of the black hole interior. 

As a somewhat different line of research, \cite{Gielen:2024lpm,Gielen:2025ovv,Gielen:2025grd} studied the interior of an AdS Schwarzschild black hole in the context of unimodular gravity ({\it \`a la} Henneaux-Teitelboim \cite{Henneaux:1989zc}), where the cosmological constant is promoted to a dynamical variable and the natural time variable conjugate to the cosmological constant emerges. By performing a self-adjoint extension of the operator in the WDW equation, either in the interior of a black hole \cite{Gielen:2024lpm} or over the entire black hole spacetime \cite{Gielen:2025ovv}, it has been shown that the black hole singularity can be resolved.

In this paper, we investigate wave packet solutions of the WDW equation for the interior of static black holes, compare their behavior with the classical trajectory, and explore the quantum effects on singularity formation. We study black holes with an arbitrary cosmological constant $\Lambda$ and constant-curvature horizons, thereby encompassing Schwarzschild, Schwarzschild-dS, Schwarzschild-AdS, and topological AdS black holes.
In Sec. \ref{section2}, the WDW equation is derived, and the general solution as well as the WKB solution to it is given. 
In Sec.~\ref{section3}, by solving the WDW equation numerically, we show that in the strong quantum-gravity regime, where the parameter corresponding to the gravitational constant $\kappa = \mathcal{O}(1)$ (defined below), the wave packet deviates from the classical solution and tends toward singularity avoidance. To quantify this tendency of the singularity avoidance, we introduce an appropriate “clock” inside the black hole and compute the time to singularity formation as a function of the gravitational constant.
Sec.~\ref{section5} is devoted to the conclusion. The details of the numerical method are described in Appendix \ref{numerics}.

\section{Wheeler-DeWitt equation for the black-hole interior}
\label{section2}

\subsection{Hamiltonian formulation}
We adopt a metric ansatz \cite{Uglum:1992nc,Gielen:2025ovv} which covers both the exterior of a static black hole and its dynamical interior:
\begin{equation}
    ds^2=\frac{N(r)^2 U(r)}{V(r)} dr^2 - \frac{V(r)}{U(r)} dt^2 + U(r)^2 d\Omega_k^2\ , 
    \label{metric}
\end{equation}
where 
$d\Omega_k^2$ is the metric for two-manifold of constant curvature $(k=+1,0,-1)$,  
 $U(r)$ and $V(r)$ are functions of the coordinate $r$, and $N(r)$ is the lapse function. 
We adopt the convention that the variables $U$ and $V$ have length dimensions, $[U]=[V]=L$. In the interior of a black hole, the radial coordinate $r$ becomes timelike and the time coordinate $t$ becomes spacelike; correspondingly the metric coefficient $f(r):=V(r)/U(r)$ is negative. Since we focus on the inside of the black hole, we will consider the case
\begin{equation}
     U(r)<0, \quad V(r)>0 \quad \Longleftrightarrow \quad f(r)<0,
\end{equation}
for which $g_{tt}=-V/U>0$ and $g_{rr}=N^2 U/V<0$, as appropriate for the black hole interior. 
 Then, the Einstein-Hilbert action with a cosmological constant $\Lambda$ reduces to
\begin{equation}
\begin{split}
    S&=\frac{1}{\kappa_4}\int d^4 x \sqrt{-g} \left(\frac{R}{2}-\Lambda\right) 
    -\frac{1}{\kappa_4}\int d^3 x \sqrt{h} K \\    
    &=\frac{1}{\kappa}  \int dr \bigg[ \frac{U'V'}{N} +N(k -\Lambda U^2)\bigg]
\end{split}
\end{equation}
where $h$ is the determinant of the induced metric on the surface of constant $r$, and $K$ is the trace of the extrinsic curvature. $\kappa_4$ is related to the four-dimensional Newton constant $G$ via $\kappa_4 = 8\pi G$. The constant $\kappa$ is the effective gravitational constant in a one-dimensional system, which is given by
$\kappa=\kappa_4/\int dt d\Omega_k$. A prime denotes $d/dr$.
The conjugate momenta of $U$ and $V$ are
\begin{equation}
    \pi_U = \frac{V'}{\kappa N}\ ,\quad 
    \pi_V=\frac{U'}{\kappa N}\ .
\end{equation}
The Hamiltonian density is 
\begin{equation}
    \mathcal{H}=\pi_U U' + \pi_V V' -\mathcal{L}
    = \kappa N \bigg[\pi_V\pi_U -\frac{1}{\kappa^2}(k-
    \Lambda U^2)
    \bigg]\,,
\end{equation}
which yields the Hamiltonian constraint $\mathcal{H} \approx 0$, and the classical dynamics is governed by this constraint.

We define the metric of the minisuperspace as
\begin{equation}
    g_{AB}=\pmat{0 & -1/2 \\ -1/2 & 0}\ ,\quad  g^{AB}=\pmat{0 & -2 \\ -2 & 0}\ ,
\end{equation}
with $A,B=U,V$. The overall coefficient of the metric is chosen so that the line element of the minisuperspace becomes the standard form of the 2-dimensional Minkowski spacetime: $ds_{MS}^2=-dUdV$. Then, the Hamiltonian constraint is written as 
\begin{equation}
    -\frac{1}{4}g^{AB}\pi_A\pi_B -\frac{1}{\kappa^2}(k-
    \Lambda U^2)=0\ .
\end{equation}

\subsection{Classical solution}
We identify the classical Schwarzschild solution within our formalism. 
The Schwarzschild metric is given by
\begin{equation}
    ds^2=\frac{dr^2}{f(r)}-c^2 f(r)dt^2 + r^2d\Omega_k^2\ ,
\end{equation}
Here, $c$ is a constant that could be eliminated by rescaling the time variable $t$, but we keep it for later convenience.
Then, the variables $U(r)$, $V(r)$, and $N(r)$ defined in Eq.(\ref{metric}) take 
%In standard coordinates, the Schwarzschild metric inside the horizon is given by setting 
\begin{equation}
   U(r)=-r\ ,\quad V(r)=-c^2rf(r)\ ,\quad N(r)=c
    \label{Schsol}
\end{equation}
where
\begin{equation}
    f(r)=-\frac{\Lambda}{3}r^2+k-\frac{\mu}{r}\ .
\end{equation}
The event horizon is located at the positive root $r_h>0$ of $f(r_h)=0$, and $f(r)$ is negative in the interior region $0<r<r_h$. The parameter $\mu$ is proportional to the black-hole mass.

The sign of $\Lambda$ controls which horizon geometries are compatible with a static black hole. For $\Lambda=0$ (asymptotically flat) and for $\Lambda>0$ (dS), only the spherical case $k=+1$ yields a static black hole, namely the Schwarzschild solution and Schwarzschild-dS solution. 
For $\Lambda<0$ (AdS), however, all three horizon geometries are allowed, giving the family of topological AdS black holes: $k=+1$ case is Schwarzschild-AdS, $k=0$ case is a planar horizon with radius $r_h=(3\mu /|\Lambda|)^{1/3}$, and the $k=-1$ case corresponds to a hyperbolic horizon.

\begin{figure}
\begin{center}
\includegraphics[scale=0.5]{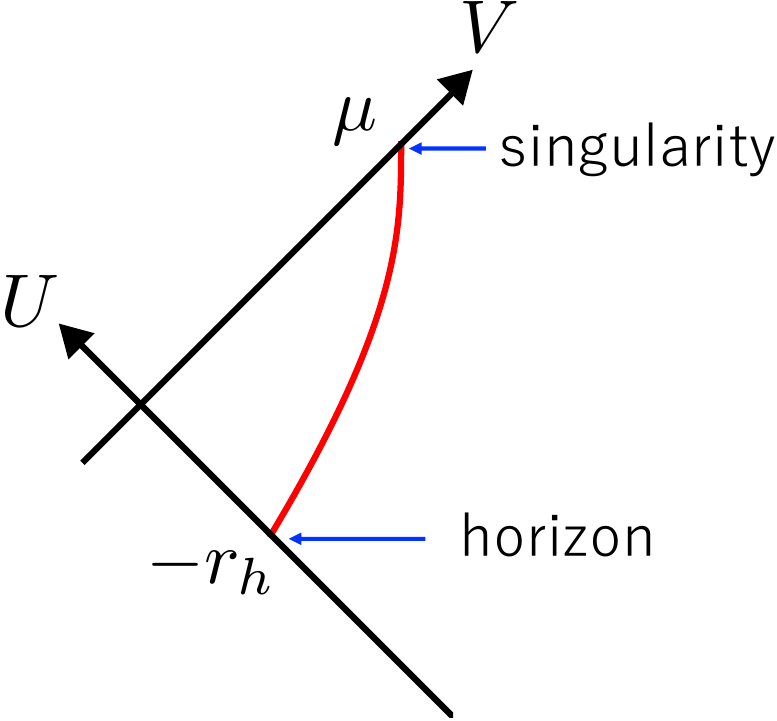}
\end{center}
\caption{Schematic picture of the classical solution in the $(U,V)$ plane. The trajectory starts at the horizon $(-r_h, 0)$ and ends at the singularity $(0, \mu)$.
}
\label{fig:classicalsol}
\end{figure}

Eliminating the parameter $r=-U$, the classical trajectory in the $(U,V)$ minisuperspace is
\begin{equation}\label{eq:classical-trajectory}
\frac{1}{c^2}V =-\frac{\Lambda}{3}U^3+kU+\mu \equiv F(U)\,,
\end{equation}
and the black hole interior runs between the singularity and the horizon, which in these variables are
\begin{equation}
(U,V)=(0,c^2\mu)\quad (r\to 0)\ ,\qquad (U,V)=(-r_h,0)\quad (r\to r_h-0)\,.
\end{equation}
A schematic picture of this classical trajectory for $c=1$ is depicted in Fig.~\ref{fig:classicalsol}.
Here, $V$-clock is "slow" in the terminology of \cite{Gotay:1983kvm}:  singularities in slow clocks are encountered in finite time. On the other hand, "fast" clocks run to $\pm\infty$ to reach singularities. For example, \cite{Hartnoll:2022snh} considered $\log(-V/U^3)$ or $\log(-U^3V)$ as a clock, which is a fast clock, and found no deviation from the classical behavior. 
According to the conjecture of \cite{Gotay:1983kvm}, one would expect singularities to be resolved quantum mechanically for a slow clock, since the unitary evolution for such a slow clock seems incompatible with the singular geometry  \cite{Gotay:1983kvm,Gielen:2024lpm,Gielen:2020abd}.

\subsection{Canonical quantization}
We proceed to canonical quantization by promoting the variables to operators acting on the wave function $\Psi(U,V)$. The Hamiltonian constraint equation $\mathcal{H}\,\Psi = 0$ becomes the WDW equation. We select the operator ordering that leads to the Laplacian in the minisuperspace metric $g_{AB}$ \cite{Hawking:1985bk} by rewriting 
\begin{equation}
    g^{AB}\pi_A \pi_B \to -(-g)^{-1/2}\partial_A (-g)^{1/2} g^{AB} \partial_B\ ,
\end{equation}
where we set $\hbar=1$. The WDW equation takes the form:
\begin{equation}
    \left[  \frac{\partial^2}{\partial U \partial V} + \frac{1}{\kappa^2}(k - \Lambda U^2) \right] \Psi(U,V) = 0\ .
    \label{eq:wdw}
\end{equation}
Equivalently, in minisuperspace coordinates $T=(V+U)/2$, $X=(V-U)/2$, we find
\begin{equation}\label{eq:wdwTX}
   \Bigl[\,\pa_T^2-\pa_X^2 + \frac{4}{\kappa^2}\bigl(k-\Lambda (T-X)^2\bigr)\Bigr]\Psi(T,X)=0\,,
\end{equation}
with $ds^2_{\rm MS}=-dT^2+dX^2$. $T$ is also a slow clock. 
Here, $\kappa$ acts as the expansion parameter for the semi-classical approximation, with the classical limit corresponding to $\kappa \to 0$. Upon restoring $\hbar$, the expansion parameter becomes $\hbar\kappa \propto \hbar G$, characterizing quantum gravitational corrections. The fact that $\kappa$ controls quantum effects can be understood more clearly from the viewpoint of the path integral. 
Since $\kappa$ appears as a prefactor of the action, it becomes a meaningless parameter if we consider only the saddle point (i.e., the classical solution). 
However, $\kappa$ controls the magnitude of quantum fluctuations around the saddle point, and therefore it affects physical phenomena in the quantum theory. 
Because our aim is to study how the wave function changes from the classical limit to the fully quantum regime, 
we treat $\kappa$ as a free parameter. Instead of fixing $\kappa$, we fix either the cosmological constant $\Lambda$ or the horizon radius $r_h$. The details of the choice of units will be explained in Sec.~\ref{section3}.

\subsection{Semi-classical solutions and interpretation}

To extract semi-classical spacetime dynamics from the WDW equation, we use the WKB approximation method. This approach is suitable when the quantum effects characterized by $\kappa$ are small compared to the scales of the black hole system. We seek semi-classical WDW solutions by introducing the WKB ansatz:
\begin{equation}
    \Psi(U,V) = A(U,V) \exp\left( \frac{i}{\kappa} S(U,V) \right)\,,
\end{equation}
where $S(U,V)$ represents the classical action and $A(U,V)$ is a slowly varying amplitude. In the leading-order approximation, we can set $A \approx \text{const}$. Substituting this into the WDW equation \eqref{eq:wdw} and expanding in powers of $\kappa$, the leading order term ($\mathcal{O}(\kappa^{-1})$) gives the Hamilton-Jacobi equation
\begin{equation}\label{eq:Hamilton-Jacobi}
 \frac{\partial S}{\partial U} \frac{\partial S}{\partial V}
 = k - \Lambda U^2\,.
\end{equation}

To solve the Hamilton-Jacobi equation~\eqref{eq:Hamilton-Jacobi}, we employ the method of separation of variables, and seek solutions of the form $S(U,V) = S_U(U) + S_V(V)$.
Substituting this ansatz into the Hamilton-Jacobi equation~\eqref{eq:Hamilton-Jacobi}, we obtain
\begin{equation}
    \frac{dS_U}{dU} \cdot \frac{dS_V}{dV} = k - \Lambda U^2\,.
\end{equation}
Since the right-hand side of the above equation is independent of $V$, we can set
\begin{equation}
    \frac{dS_V}{dV} = -\frac{1}{c}\,, \quad \frac{dS_U}{dU} = -c(k - \Lambda U^2)\,,
\end{equation}
where $c$ is a constant. 
Integrating these equations, we obtain
\begin{equation}
    S(U,V) = S_U(U)+S_V(V)= -c \left(-\frac{\Lambda}{3} U^3 + kU\right) - \frac{V}{c}+ \text{const}\,,
\end{equation}
% \begin{equation}
%     S_V(V) = -\frac{V}{c} + \text{const}\,, \quad S_U(U) 
%     = -c\left(-\frac{\Lambda}{3} U^3 + kU\right)+ \text{const}\,,
% \end{equation}
% where $F(U)$ is given by Eq.~\eqref{eq:classical-trajectory}.
%Thus, the general solution of Eq.~\eqref{eq:Hamilton-Jacobi} becomes
%where $d$ is an integration constant that can be absorbed into the overall phase. 
Thus, the semiclassical solution is given by
\begin{equation}
    \Psi \sim \exp\left[ -\frac{i}{\kappa} \left\{
    c \left(-\frac{\Lambda}{3} U^3 + kU\right) + \frac{V}{c}
   \right\}\right]\,.
    \label{solution:wkb}
\end{equation}
In fact, (\ref{solution:wkb}) is the exact solution to the WDW equation Eq.(\ref{eq:wdw}) \cite{Uglum:1992nc,Gielen:2025ovv}. 
The classical trajectory corresponding to $S(U,V)$ obtained above is given by
\begin{equation}
\pi_U = \frac{V'}{\kappa N} = \partial_U \left(\frac{S}{\kappa}\right) = -\frac{c}{\kappa} (k - \Lambda U^2), \quad
\pi_V = \frac{U'}{\kappa N} = \partial_V \left(\frac{S}{\kappa}\right) = -\frac{1}{c\kappa}\ .
\end{equation}
From these relations, we obtain $dV/dU = c^2 (k - \Lambda U^2)$, and the classical trajectory takes the form
\begin{equation}
\frac{1}{c^2} V = -\frac{\Lambda}{3} U^3 + kU + \mu\ ,
\end{equation}
where $\mu$ is an integration constant.
This result coincides with Eq.(\ref{eq:classical-trajectory}), indicating that the constants $c$ and $\mu$ correspond to the freedom of rescaling the time variable and the mass parameter. In the following, we consider the case $c = 1$, which corresponds to the standard scaling of time.

\subsection{Retarded Green function and general solution}
We can express the general solution of the WDW equation Eq.(\ref{eq:wdw}) by a Green function $G(U,V;U',V')$ which satisfies
\begin{equation}
    \left[  \frac{\partial^2}{\partial U \partial V} + \frac{1}{\kappa^2}(k - \Lambda U^2) \right]G(U,V;U',V')=\delta(U-U')\delta(V-V')\,.
    \label{eq:green-f}
\end{equation}
The retarded Green function is given by \cite{Uglum:1992nc}
\begin{equation}
G(U,V;U',V')=\theta(U-U')\theta(V-V')J_0\left(\frac{2}{\kappa}\sqrt{(V-V')(F(U)-F(U'))}\right)\,,
\end{equation}
where $J_0$ is the Bessel function of the first kind of order zero. When $F(U)-F(U')<0$, we interpret $J_0(2\kappa^{-1}\sqrt{(V-V')(F(U)-F(U'))})=I_0(2\kappa^{-1}\sqrt{(V-V')(F(U')-F(U)}))$ with $I_{0}$ being the modified Bessel function of the first kind. Given the boundary data at $U=U_B,V=V_B$, the general solution is then written as
\begin{multline}
    \Psi(U,V)=\frac{1}{2}\int^\infty_{U_B}dU' G(U,V;U',V_B) \overleftrightarrow{\partial}_{U'}\Psi(U',V_B)\\
    +\frac{1}{2}\int^\infty_{V_B}dV' G(U,V;U_B,V') \overleftrightarrow{\partial}_{V'}\Psi(U_B,V')\ 
    \label{sol:wdw-green}
\end{multline}
where $f \overleftrightarrow{\partial}_A g = f\partial_A g - g \partial_A f$ ($A=U,V$).
By performing the above integral on the initial surface, one can obtain a solution to the WDW equation for arbitrary initial data.

\section{Solutions of Wheeler-DeWitt equation}
\label{section3}

To study the qualitative behavior of the wave function, we numerically solve the WDW equation. 
If Eq.(\ref{sol:wdw-green}) is used directly, the integral must be evaluated each time the wave function is computed at a point in spacetime, which is computationally costly. Therefore, in practice, we discretize the null-coordinate $(U,V)$ and construct the numerical solution using a finite-difference method. The details of the numerical calculation are provided in Appendix.\ref{numerics}.

Up to this point, we have adopted the unit system $c=\hbar=1$, and there remains the freedom to fix one more dimensional quantity. In this paper, for cases that are asymptotically dS or AdS, we choose a unit system in which the curvature radius is set to unity; that is, we set $|\Lambda|=3$. On the other hand, for asymptotically flat spacetimes, we fix the horizon radius. Here, in order to facilitate comparison with the dS and AdS cases, we set $r_h=0.3$.

Let us consider the initial condition for the wave function. 
In principle, the initial condition is not determined \textit{a priori}. 
Therefore, we assume that the initial data are identical to the semiclassical solution. 
This assumption ensures that the wave function is localized in the near-horizon regime ($V \simeq 0$), as in the classical theory. 
Our aim is then to investigate how quantum effects influence the subsequent dynamics of the wave function.
We take the initial surface at $V=0$ and $U=U_B$. 
%To capture the quantum state of a black hole in the WDW equation, we need to specify an appropriate initial state at $V=V_B = 0$.
A suitable choice is a Gaussian wave packet modulated by the classical phase:
\begin{equation}
    \Psi(U, V=0) = \exp\left[-\frac{(U - U_0)^2}{2\sigma^2}\right] \exp\left[-\frac{i}{\kappa} F(U)\right]\,,\quad
    \Psi(U_B, V) = 0\ ,
    \label{eq:initial-state}
\end{equation}
where $U_0 = -r_h$ corresponds to the horizon radius, the dispersion $\sigma$ characterizes the quantum uncertainty,  
and the phase factor $\exp[-i F(U)/\kappa]$ ensures the correct classical evolution. $U_B$ is chosen such that the Gaussian of the initial data becomes negligibly small at $U = U_B$  (for example, $U_B \sim U_0 - 5\sigma$).
This initial state~\eqref{eq:initial-state} represents a quantum black hole with quantum fluctuations controlled by $\sigma$. The limit $\sigma \to 0$, $\kappa \to 0$, with fixing $\kappa/\sigma\ll 1$, corresponds to the classical trajectory.

\begin{figure}[t]
\begin{minipage}[t]{0.31\linewidth}
    \centering
    \includegraphics[width=\linewidth]{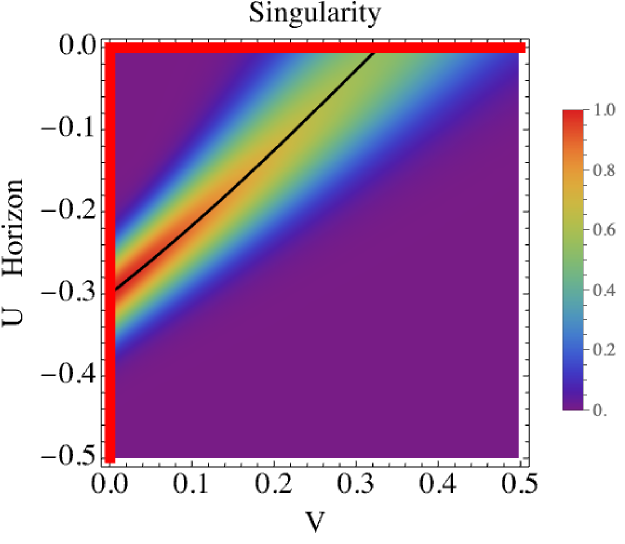}
    \subcaption{$\Lambda=-3$ and $\kappa=0.01$}
  \end{minipage}\hspace{0.02\linewidth}
  \begin{minipage}[t]{0.31\linewidth}
    \centering
    \includegraphics[width=\linewidth]{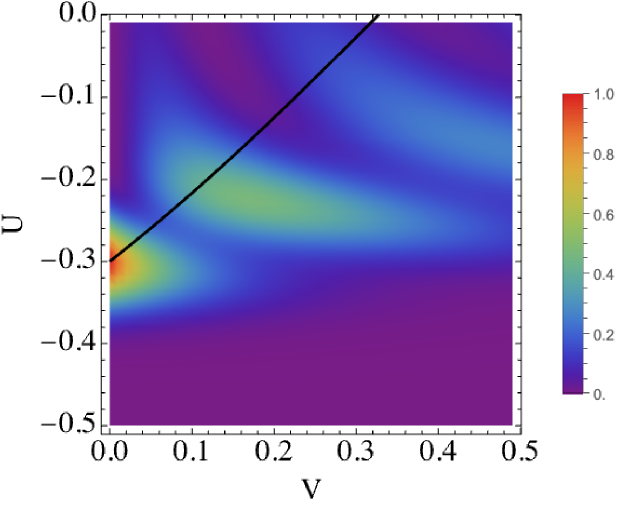}
    \subcaption{$\Lambda=-3$ and $\kappa=0.1$}
  \end{minipage}
  \hspace{0.02\linewidth}
    \begin{minipage}[t]{0.31\linewidth}
    \centering
    \includegraphics[width=\linewidth]{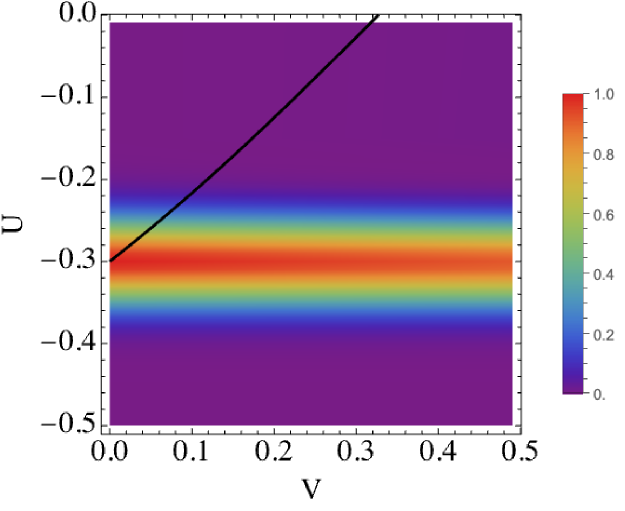}
    \subcaption{$\Lambda=-3$ and $\kappa=1$}
  \end{minipage}\vspace{0.5em}
%%%%%%%%%%%%%%%%%%%%%%%%%%%%%%%%%%%%%%%%%
   \begin{minipage}[t]{0.31\linewidth}
    \centering
    \includegraphics[width=\linewidth]{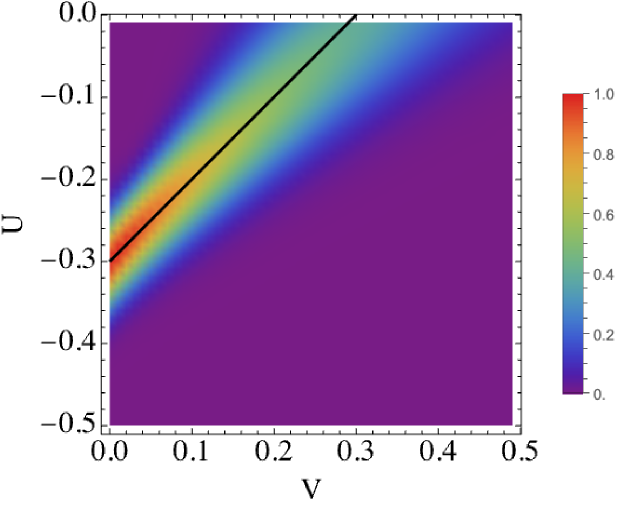}
    \subcaption{$\Lambda=0$ and $\kappa=0.01$}
  \end{minipage}
  \hspace{0.02\linewidth}
  \begin{minipage}[t]{0.31\linewidth}
    \centering
    \includegraphics[width=\linewidth]{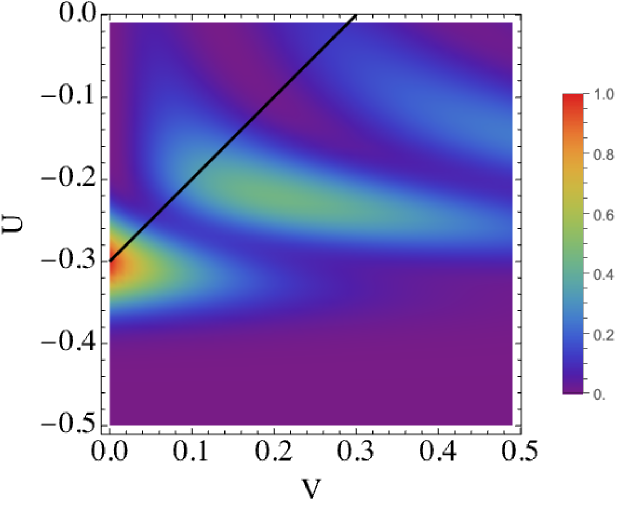}
    \subcaption{$\Lambda=0$ and $\kappa=0.1$}
  \end{minipage}
  \hspace{0.02\linewidth}
\begin{minipage}[t]{0.31\linewidth}
    \centering
    \includegraphics[width=\linewidth]{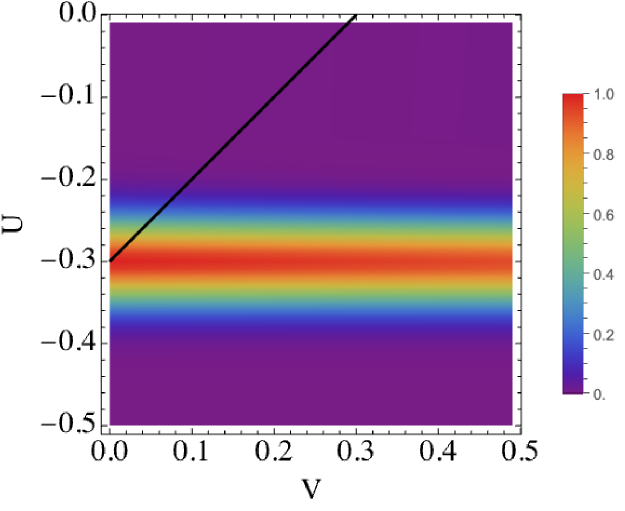}
    \subcaption{$\Lambda=0$ and $\kappa=1$}
  \end{minipage}\vspace{0.5em}
%%%%%%%%%%%%%%%%%%%%%%%%%%%%%%%%%%%%%%%%%
  \begin{minipage}[t]{0.31\linewidth}
    \centering
    \includegraphics[width=\linewidth]{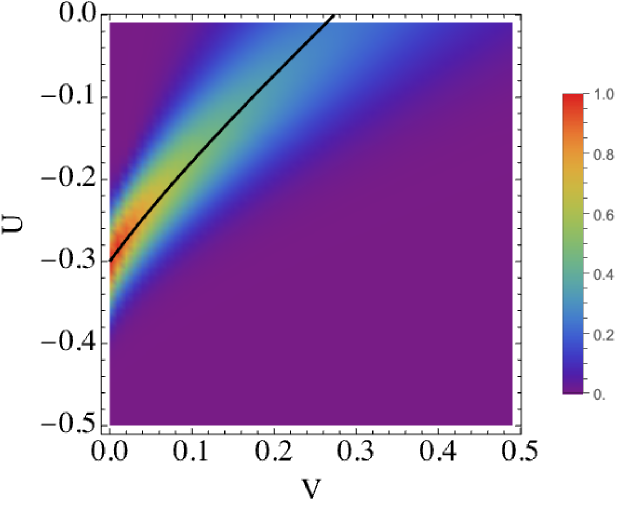}
    \subcaption{$\Lambda=3$ and $\kappa=0.01$}
  \end{minipage}
  \hspace{0.02\linewidth}
  \begin{minipage}[t]{0.31\linewidth}
    \centering
    \includegraphics[width=\linewidth]{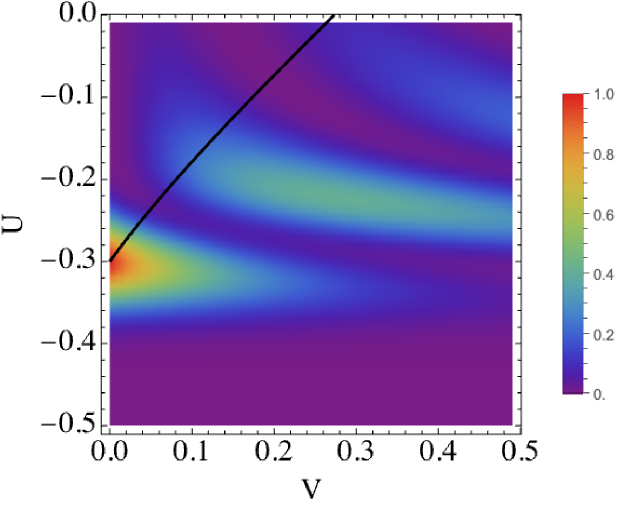}
    \subcaption{$\Lambda=3$ and $\kappa=0.1$}
  \end{minipage}
  \hspace{0.02\linewidth}
\begin{minipage}[t]{0.31\linewidth}
    \centering
    \includegraphics[width=\linewidth]{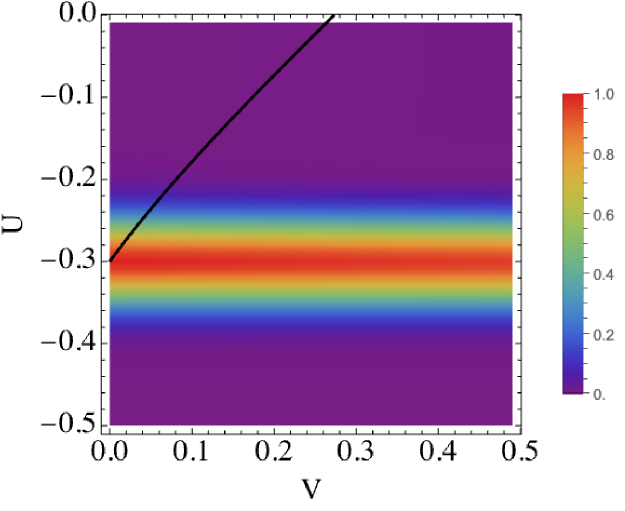}
    \subcaption{$\Lambda=3$ and $\kappa=1$}
  \end{minipage}
  \caption{
  Squared wave function $|\Psi(U,V)|^2$ for spatially closed geometries ($k=+1$) with $\Lambda=-3$ (upper), $0$ (middle), and $3$ (lower) for several values of $\kappa$. These correspond to the Schwarzschild-AdS, Schwarzschild, and Schwarzschild-dS black holes. The Gaussian width in the initial state~\eqref{eq:initial-state} is fixed to $\sigma = 0.05$, and $r_h = 0.3$ is chosen. The initial data are chosen such that the semiclassical solution and the wave function coincide in the near-horizon regime. The boundary \(U_B\) is set sufficiently far from the center of the wave packet ($U_B = -0.5$), so that the initial data vanish there.
The black curve represents the classical trajectory, which reaches the black hole singularity at $U=0$.  
  }
  \label{fig:closed-kappa-variable}
\end{figure}

Figure~\ref{fig:closed-kappa-variable} shows the square of the modulus of the wave function $|\Psi(U,V)|^2$ for spatially closed geometries ($k = +1$) with varying quantum effects $\kappa$. Each row corresponds to Schwarzschild-AdS ($\Lambda = -3$), Schwarzschild ($\Lambda = 0$), and Schwarzschild-dS ($\Lambda = +3$) black holes, while columns correspond to increasing the quantum gravitational effects $\kappa = 0.01, 0.1, 1$.

%%%%%%%%%%%%%%%%%%%%%%%%%%%%%%%%%%%%%%%%%
%%%%%%%%%%%%%%%%%%%%%%%%%%%%%%%%%%%%%%%%%
%%%%%%%%%%%%%%%%%%%%%%%%%%%%%%%%%%%%%%%%%
%%%%%%%%%%%%%%%%%%%%%%%%%%%%%%%%%%%%%%%%%

In the semi-classical regime ($\kappa = 0.01$), the wave packet closely follows the classical trajectory Eq. (\ref{eq:classical-trajectory}) shown by the black curve, maintaining a narrow width throughout its evolution toward the singularity at $U = 0$. 
The smallness of this spreading supports the validity of the semi-classical approximation. As $\kappa$ increases to $0.1$, quantum effects become noticeable through increased wave packet dispersion, though the evolution still follows the classical path. In the strong quantum regime ($\kappa = 1$), the wave packet deviates significantly from the classical trajectory and appears to go straight along the $U=$constant line, 
suggesting a quantum resolution of the singularity.
Since the potential term of the WDW equation is proportional to $1/\kappa^2$, 
the effect of scattering due to the potential on the wave packet becomes weaker as $\kappa$ increases.  
Thus, for large $\kappa$, the wave packet propagates linearly along constant-$U$ surfaces.  
Because $U=0$ corresponds to the singularity, this suggests that strong quantum effects tend to make the wave packet avoid the singularity.  
In the next section, we compute the expectation value of the metric component and quantitatively evaluate the time required for singularity formation as a function of $\kappa$.

\begin{figure}[t]
\begin{minipage}[t]{0.31\linewidth}
    \centering
    \includegraphics[width=\linewidth]{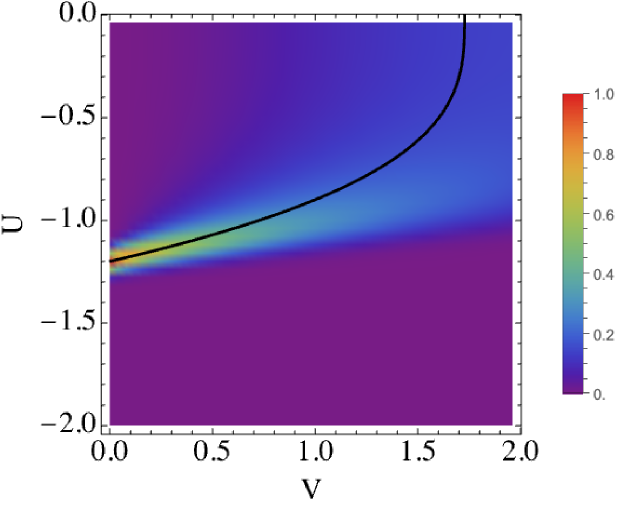}
    \subcaption{$\kappa=0.1$}
  \end{minipage}\hspace{0.02\linewidth}
  \begin{minipage}[t]{0.31\linewidth}
    \centering
    \includegraphics[width=\linewidth]{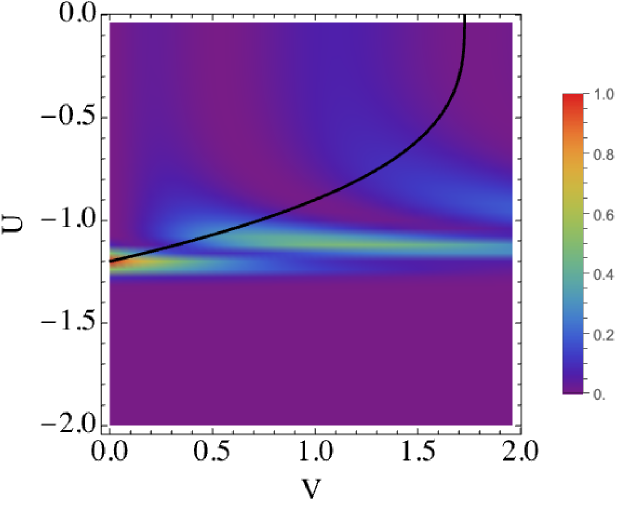}
    \subcaption{$\kappa=0.5$}
  \end{minipage}
  \hspace{0.02\linewidth}
    \begin{minipage}[t]{0.31\linewidth}
    \centering
    \includegraphics[width=\linewidth]{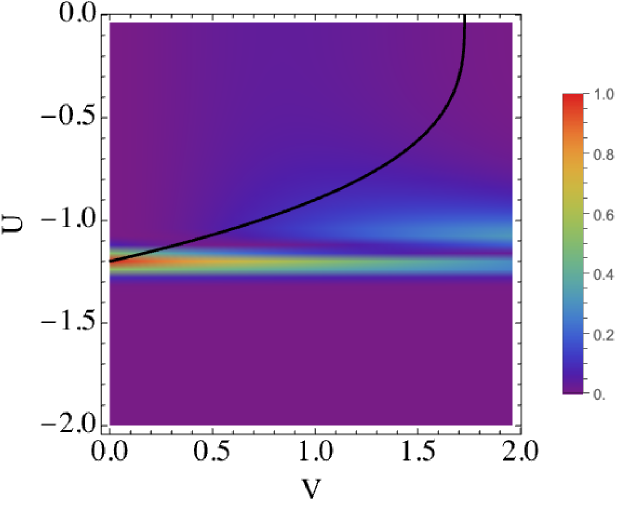}
    \subcaption{$\kappa=1$}
  \end{minipage}\vspace{0.5em}
%%%%%%%%%%%%%%%%%%%%%%%%%%%%%%%%%%%%%%%%%
   \begin{minipage}[t]{0.31\linewidth}
    \centering
    \includegraphics[width=\linewidth]{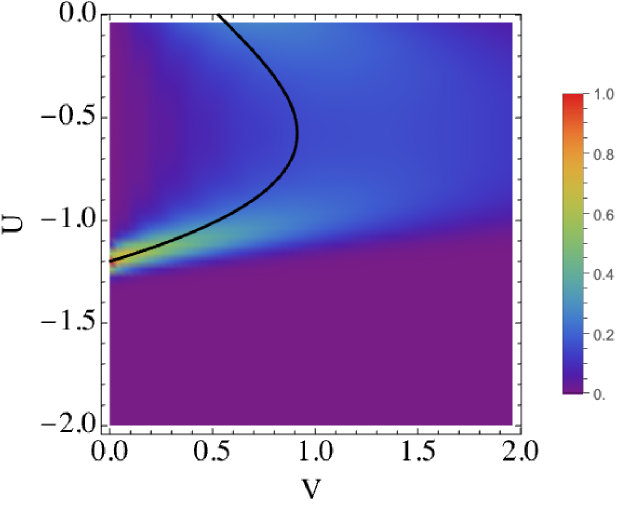}
    \subcaption{$\kappa=0.1$}
  \end{minipage}
  \hspace{0.02\linewidth}
  \begin{minipage}[t]{0.31\linewidth}
    \centering
    \includegraphics[width=\linewidth]{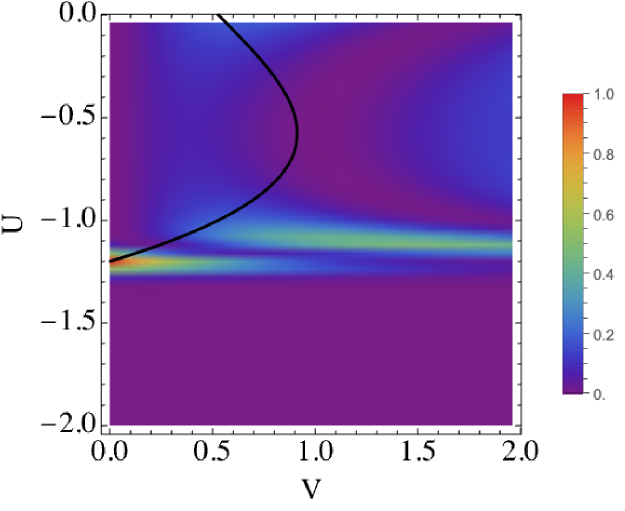}
    \subcaption{$\kappa=0.5$}
  \end{minipage}
  \hspace{0.02\linewidth}
\begin{minipage}[t]{0.31\linewidth}
    \centering
    \includegraphics[width=\linewidth]{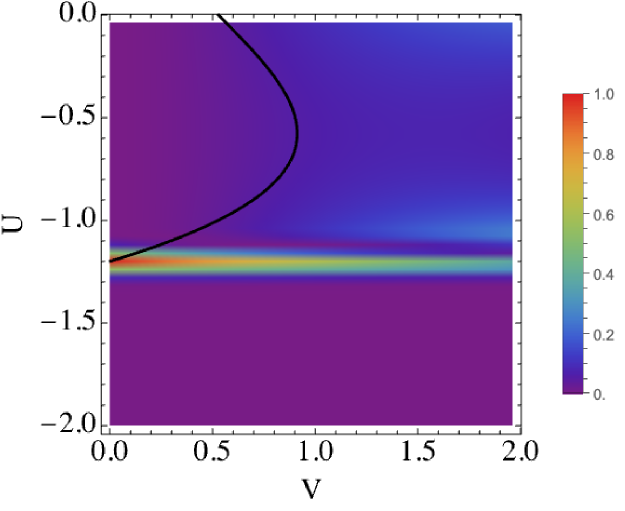}
    \subcaption{$\kappa=1$}
  \end{minipage}
%%%%%%%%%%%%%%%%%%%%%%%%%%%%%%%%%%%%%%%%%
  \caption{
  Squared wave function $|\Psi(U,V)|^2$ for spatially flat ($k=0$, upper) and open ($k=-1$, lower) geometries with a negative cosmological constant $\Lambda = -3$ for several values of $\kappa$.
These correspond to topological AdS black holes.
The Gaussian width in the initial state~\eqref{eq:initial-state} is fixed to $\sigma = 0.05$, and $r_h = 1.2$ is chosen. We set $U_B =-2.0$.
}
  \label{fig:flat-open-kappa-variable}
\end{figure}

Figure~\ref{fig:flat-open-kappa-variable} show $|\Psi(U,V)|^2$ of topological AdS black holes ($\Lambda = -3$) with flat ($k = 0$) and hyperbolic ($k = -1$) spatial geometries. These exotic black hole solutions, which exist only in the presence of a negative cosmological constant, exhibit different quantum behaviors dependent on their spatial topology.
As $\kappa$ increases from $0.1$ to $1$, both topologies show the same behaviors with the Schwarzschild black holes. In the intermediate ($\kappa = 0.5$) and strong ($\kappa = 1$) quantum regimes, $|\Psi(U,V)|^2$ near $U = 0$ becomes significantly suppressed, indicating quantum resolution of the singularity. 

Depending on the choice of parameters, solving the WDW equation can sometimes lead to difficulties.  The WDW equation can be regarded as a Klein-Gordon equation with a potential term, where the potential is given by $\mathcal{V} = 4(k - \Lambda U^2)/\kappa^2$. This potential is positive definite when $k \geq 0$ and $\Lambda \leq 0$, but it can take negative values otherwise. In the region where $\mathcal{V} < 0$, a so-called tachyonic instability arises, and the wave function grows exponentially along the "time" direction. Such behavior has indeed been observed in numerical computations. Although the physical interpretation of this instability remains unclear, we restricted our discussion to the region where the tachyonic instability does not play a significant role.

%%%%%%%%%%%%%%%%%%%%%%%%%%%%%%%%%%%%%%%%%
%%%%%%%%%%%%%%%%%%%%%%%%%%%%%%%%%%%%%%%%%
%%%%%%%%%%%%%%%%%%%%%%%%%%%%%%%%%%%%%%%%%
%%%%%%%%%%%%%%%%%%%%%%%%%%%%%%%%%%%%%%%%%

\section{Choice of the clock and expectation value}
\label{section4}

%%%%%%%%%%%%%%%%%%%%%%%%%%%%%%%%%%%%%%%%%
%%%%%%%%%%%%%%%%%%%%%%%%%%%%%%%%%%%%%%%%%
\begin{figure}[t]
\begin{minipage}[t]{0.31\linewidth}
    \centering
    \includegraphics[width=\linewidth]{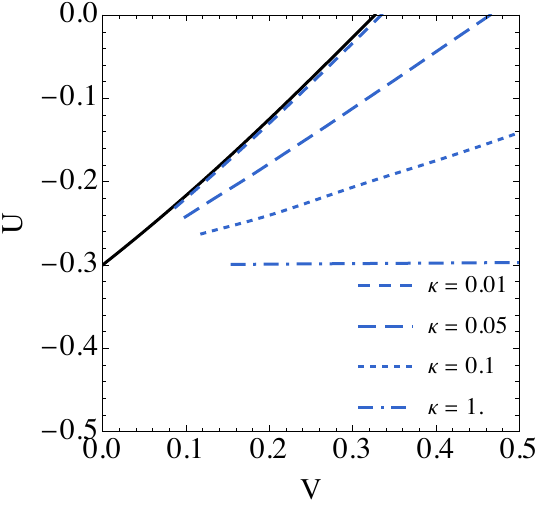}
    \subcaption{Schwarzschild-AdS}
  \end{minipage}\hspace{0.02\linewidth}
  \begin{minipage}[t]{0.31\linewidth}
    \centering
    \includegraphics[width=\linewidth]{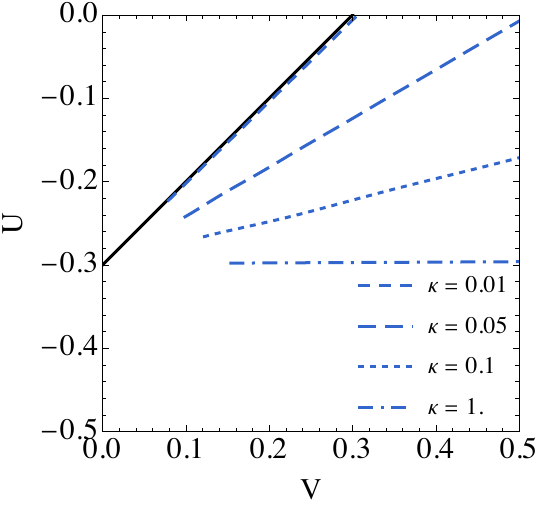}
    \subcaption{Schwarzschild}
  \end{minipage}
  \hspace{0.02\linewidth}
    \begin{minipage}[t]{0.31\linewidth}
    \centering
    \includegraphics[width=\linewidth]{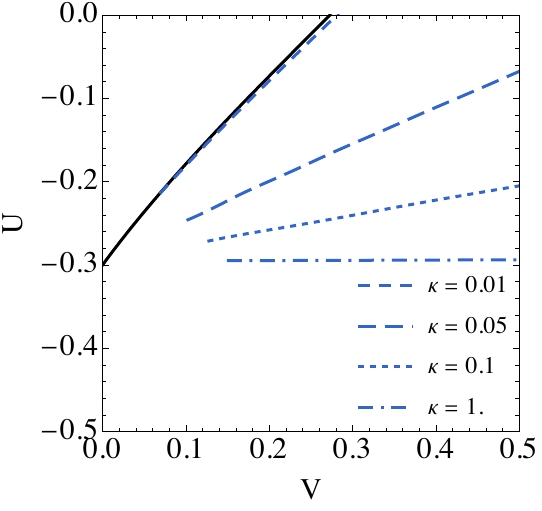}
    \subcaption{Schwarzschild-dS}\bigskip
  \end{minipage} 
%%%%%%%%%%%%%%%%%%%%%%%%%%%%%%%%%%%%%%%%%
\centering
\begin{minipage}[t]{0.31\linewidth}
    \centering
    \includegraphics[width=\linewidth]{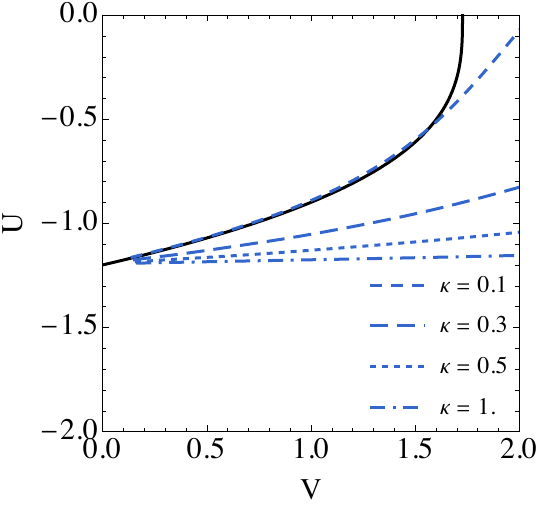}
    \subcaption{Planar-AdS}
  \end{minipage} \hspace{0.02\linewidth}
    \begin{minipage}[t]{0.31\linewidth}
    \centering
    \includegraphics[width=\linewidth]{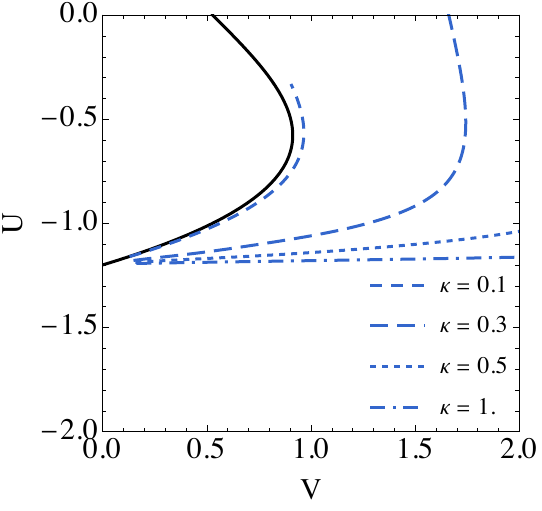}
    \subcaption{Hyperbolic-AdS}
  \end{minipage}
%%%%%%%%%%%%%%%%%%%%%%%%%%%%%%%%%%%%%%%%%
  \caption{The expectation value $\langle X\rangle$ for Schwarzschild black holes with cosmological constant $\Lambda\in\{-3,0,+3\}$ and for topological AdS black holes with $k\in\{0,-1\}$, as a function of $\kappa$. The figure is plotted in the $(V, U)$ plane, using $V=T+X$ and $U=T-X$. We adopt the same parameter choices for $\sigma$ and $r_h$ as in Figures~\ref{fig:closed-kappa-variable} and~\ref{fig:flat-open-kappa-variable}. The black curve corresponds to the classical trajectory, while the blue curves correspond to the expectation value.  Blue curves near $V=0$ are missing because data interpolation was not performed properly during the transformation to the $(V, U)$ plane. See the text for the details.
  }
  \label{fig:xvev-kappa-variable}
\end{figure}
%%%%%%%%%%%%%%%%%%%%%%%%%%%%%%%%%%%%%%%%%
%%%%%%%%%%%%%%%%%%%%%%%%%%%%%%%%%%%%%%%%%

The covariant form of the WDW equation in minisuperspace is written as
\begin{equation}
    [\Box -\frac{4}{\kappa^2}(k-\Lambda U^2)]\Psi=0\ .
\end{equation}
Let us assume that both $\Psi_1$ and $\Psi_2$ satisfy the WDW equation.  
On a hypersurface $\Sigma$, we define the inner product for wave functions as  
\begin{equation}
    (\Psi_1, \Psi_2) = -i \int_\Sigma d\Sigma_A  (\Psi_1^\ast \partial^A \Psi_2-\Psi_2 \partial^A \Psi_1^\ast) \ . \label{inprod_covariant}
\end{equation}  
Here, we follow the notation of \cite{Poisson:2009pwt} for the surface element $d\Sigma_A$. When the boundary terms can be neglected, the WDW equation implies that the inner product is independent of the choice of hypersurface. 
To neglect the boundary term, at least two possible approaches can be considered. 
The first is to restrict the Hilbert space of wave functions, as done in Ref.~\cite{Gielen:2024lpm}, and to consider only those wave functions for which the inner product is conserved. (In the context of our formulation, this corresponds to imposing a restriction on the initial data at the initial surface $V = 0$, $U = U_B$.) The second approach is to extend the domain of the minisuperspace~\cite{Hartnoll:2022snh}. As can be seen from the WDW equation above, the equation itself remains regular even at the singularity $U = 0$. Therefore, the solutions of the WDW equation can be extended over the entire spacetime ($-\infty < U, V < \infty$). In this case, the integration region of the inner product can be extended to spatial infinity, so that the boundary term can be neglected. In this work, we adopt the latter approach.

Assuming that the black hole interior is foliated by hypersurfaces $\Sigma$, this foliation defines a notion of "time". There is an arbitrariness in the choice of time, and depending on this choice, the expectation values of physical observables may vary~\cite{Hartnoll:2022snh,Gielen:2020abd,Gotay:1983kvm}. 
By introducing new coordinates in the minisuperspace defined by $T = (V + U)/2$ and $X = (V - U)/2$, the metric takes the form $ds_{\text{MS}} = -dT^2 + dX^2$.
In what follows, we consider hypersurfaces $\Sigma$ defined by constant $T$, and regard $T$ as the ``time'' parameter, to study the time evolution of expectation values of physical quantities.\footnote{
As an alternative choice, one might consider taking the $V = \text{const}$ surface as the hypersurface $\Sigma$, treating $V$ as the clock. In that case, the inner product is given by
\[
(\Psi_1,\Psi_2)=i\int_{-\infty}^\infty dU \left( \Psi_1^\ast \partial_U \Psi_2 - \Psi_2 \partial_U \Psi_1^\ast \right).
\]
However, using this inner product, we generally have
$(\Psi_1, U \Psi_2) \neq (U \Psi_1, \Psi_2)$,
which implies that the operator $U$ is not Hermitian. Consequently, the expectation value of $U$ becomes a complex, making its physical interpretation ambiguous. For this reason, we adopt Cartesian coordinates $(T, X)$ and treat $T$ as the time coordinate in this work. As discussed below Eq.(\ref{inprod_covariant}), one might think that $U$ could be made Hermitian by imposing suitable boundary conditions on the wave function. 
However, since it is straightforward to verify that 
$(U\Psi, \Psi) - (\Psi, U\Psi) \propto \int dU\, |\Psi|^2$, 
$U$ is non-Hermitian regardless of the boundary conditions, as long as a nontrivial wave function is considered.
} Then, the inner product is written as
\begin{equation}
    (\Psi_1, \Psi_2) = i \int_{-\infty}^\infty  dX  (\Psi_1^\ast \partial_T \Psi_2-\Psi_2 \partial_T \Psi_1^\ast)\ . \label{inprod}
\end{equation}
In general, the wave function leaks into the region with \(U>0\), so if one computes the inner product only over the region with \(U<0\), it is no longer conserved. Therefore, as explained in the previous paragraph in actual calculations, the computational domain of the wave function is extended into the \(U>0\) region, and the solutions in that region are also included in the evaluation of the inner product. The "norm" defined by this inner product is not, in general, positive definite. For example, for semiclassical solutions, $\Psi=A(U,V)\exp[-i (F(U)+V)/\kappa]$, we have
\begin{equation}
    (\Psi, \Psi) \simeq \frac{2}{\kappa} \int_{-\infty}^\infty dX (1+k-\Lambda U^2) |\Psi|^2\ ,
\end{equation}
at the leading order of $1/\kappa$. For $\Lambda\leq 0$ and $k\geq 0$, the above expression is positive definite, but can be negative for other cases. Even if the conditions \(\Lambda \leq 0\) and \(k \geq 0\) are satisfied, the norm can still become negative in the fully quantum regime. 
Leaving aside the issue of negative norms, we apply a naive probabilistic interpretation using the inner product introduced above, and investigate the expectation values of physical observables. In particular, the expectation value of the metric component $X$ is defined as 
\begin{equation}
    \langle X \rangle = \frac{(\Psi, X \Psi)}{(\Psi,\Psi)}\,, 
\end{equation}
where $(\Psi, X \Psi)$ is written as 
\begin{equation}
  (\Psi, X\Psi) = i \int_{-\infty}^\infty  dX\, X  (\Psi^\ast \partial_T \Psi-\Psi \partial_T \Psi^\ast)\,.
  \label{psiXpsi}
\end{equation}

\begin{figure}
\begin{center}
\includegraphics[scale=1.0]{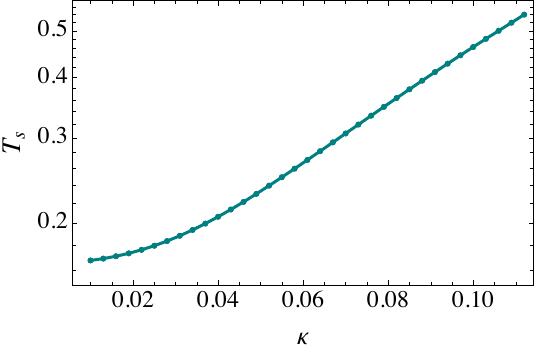}
\end{center}
\caption{
Time $T_s$ to reach the singularity (at $U=0$) as a function of $\kappa$ for the Schwarzschild–AdS black hole with $\Lambda = -3$.
$T_s$ is plotted on a logarithmic scale.
}
\label{fig:time-singularity}
\end{figure}

In Fig. \ref{fig:xvev-kappa-variable}, we show the expectation value $\langle X\rangle$ for each black hole for several $\kappa$.\footnote{
Due to limitations of the numerical method, $\langle X \rangle$ cannot be 
computed at the early stage of the time evolution. Consequently, the curve 
representing $\langle X \rangle$ in this figure is truncated near its 
initial point. For further details, see Appendix \ref{numerics}.
} 
We find that as $\kappa$ increases and the quantum effects become strong, $\langle X\rangle$ deviates from the classical trajectory (black curve).  
In fact, as shown in Fig. \ref{fig:time-singularity}, the time $T_s$ to reach the singularity $U=0$ for the Schwarzschild-AdS black hole becomes longer as $\kappa$ increases and tends to grow exponentially as $\kappa$ becomes large. $T_s$ is fitted as $T_s\propto \exp(0.4\kappa/r_h)$ around $\kappa\sim 0.1$. 
Similar behavior is found for other black holes.
See Fig. \ref{fig:time-singularity-appendix} in Appendix \ref{singularity-time} for the details.

\section{Conclusion}
\label{section5}

We have quantized the interior of a static spherically symmetric black hole with or without a cosmological constant via the WDW equation.  
We have solved the WDW equation numerically to study the behavior of a wave packet. 
Through the use of a "slow" clock, we have found that in the semiclassical regime ($\kappa\ll 1$) the wave packet follows the classical trajectory, while for the strong quantum-gravity regime ($\kappa=O(1)$) the wave packet deviates from the classical trajectory and the time to reach the classical singularity tends to diverge, which suggests
the avoidance of singularity. 
This avoidance of the singularity appears robust against the sign of the cosmological constant and horizon topology. We calculated the time $T_s$ required for the formation of the singularity and found that stronger quantum effects lead to a delay in the formation of the singularity.

It would be interesting to extend our analysis to more general systems. For instance, considering a charged black hole introduces additional degrees of freedom associated with the electromagnetic field, and one could investigate the solutions of the WDW equation, including these variables (recent studies in this direction include \cite{Blacker:2023ezy,Chien:2025tzm}). A charged black hole possesses an inner horizon, which corresponds to a Cauchy horizon, beyond which classical general relativity loses its predictability. It would therefore be intriguing to explore whether the formation of such a Cauchy horizon can be suppressed by quantum gravitational effects.

Furthermore, in the case of odd-dimensional rotating black holes with all angular momenta set equal, the metric depends only on a single radial coordinate, making the spacetime cohomogeneity-1~\cite{Gibbons:2004uw}. In this situation, the minisuperspace approximation employed in the present work could also be applicable, allowing us to analyze the WDW equation inside the black hole. In particular, it would be interesting to study how quantum effects influence the Cauchy horizon that arises due to rotation.
We hope to report the results of the investigation in the future.

\section*{Acknowledgments}
%%%%%%%%%%%%%%%%%%%%%%%%%%%%%%%%%%%%%
This work is supported by JSPS Grant-in-Aid for Scientific Research Number
22K03640 (TC), 23K13100 (HM), JP21H05186 (KM) and in part by Nihon University.

\appendix

\section{Numerical method}
\label{numerics}

In order to solve the WDW equation~(\ref{eq:wdw}),
rather than evaluating the integrals in Eq.~(\ref{sol:wdw-green}), which is numerically demanding,
we directly solve the WDW equation~(\ref{eq:wdw}) using the finite-difference method.

\begin{figure}
\begin{center}
\includegraphics[scale=0.7]{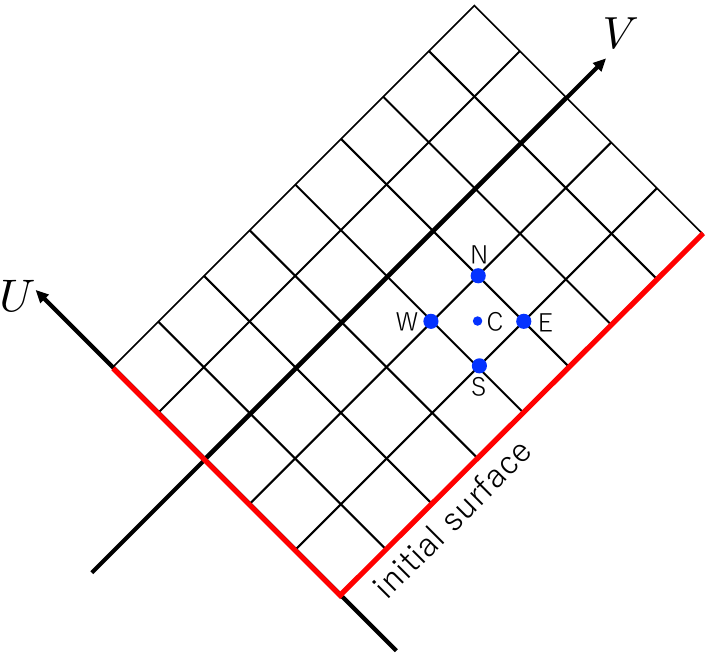}
\end{center}
\caption{Domain for the numerical integration of the WDW equation. 
}
\label{fig:Ndomain}
\end{figure}

As shown in Eq.~(\ref{solution:wkb}), in the vicinity of the classical limit ($\kappa \to 0$), the phase of the wave function,
$e^{-i(F(U)+V)/\kappa}$, oscillates rapidly. When treating this numerically, the mesh size must be much smaller than the corresponding wavelength, which makes the computation highly costly. To address this issue, we define a new function by factoring out the rapidly oscillating phase as
\begin{equation}
    \Psi(U,V)=\exp[-\frac{i}{\kappa}(F(U)+V)]\psi(U,V)\ .
\end{equation}
Then, the WDW equation is rewritten as
\begin{equation}
    \partial_U \partial_V \psi -\frac{i}{\kappa}(\partial_U\psi + F'(U)\partial_V\psi)=0\ .
    \label{psieq}
\end{equation}
As shown in Fig.\ref{fig:Ndomain}, we discretize the $(U,V)$-coordinates with grid spacing $h$.
Let us focus on the points $N$, $E$, $W$, $S$, and $C$ indicated in the figure.  
Using the values of the wave function at the points $N$, $E$, $W$, and $S$, the derivative of the wave function at the point $C$ can be evaluated as
\begin{equation}
\begin{split}
    &\partial_U\psi|_C=\frac{\psi|_N-\psi|_E+\psi|_W-\psi|_S}{2h}\ ,\quad 
    \partial_V\psi|_C=\frac{\psi|_N-\psi|_W+\psi|_E-\psi|_S}{2h}\ ,\\
    &\partial_U\partial_V\psi|_C=\frac{\psi|_N-\psi|_E-\psi|_W+\psi|_S}{h^2}\ .
    \end{split}
\end{equation}
By substituting the above into Eq.(\ref{psieq}) and solving for $\psi|_N$, we obtain
\begin{equation}
    \psi|_N=\frac{\{1-\alpha(F'-1)\}\psi|_W+\{1+\alpha(F'-1)\}\psi|_E-\{1+\alpha (F'+1)\}\psi|_S}{1-\alpha (F'+1)}
\end{equation}
where $\alpha=ih/(2\kappa)$.
Using the above equation, one can generate the data at $N$ from the numerical data at $W$, $E$, and $S$.  
Since the initial data are provided on the initial surface in Fig.\ref{fig:Ndomain} by Eq.(\ref{eq:initial-state}), successive application of this equation yields the data in the numerical domain shown in the figure.

In this paper, we compute the expectation value of $X = (V - U)/2$ from the wave function.
For this purpose, it is necessary to perform the numerical integration~(\ref{psiXpsi}) along surfaces of constant time $T = (U + V)/2$.  That is, we make use of the data on the mesh shown in Fig.\ref{fig:Ndomain} in the horizontal direction.  
However, since we are considering a rectangular region in the $(U,V)$ plane, for small or large values of $T$, the constant-$T$ surface only grazes the corners of the rectangular domain,  
so that sufficient numerical data cannot be obtained in those time regions. 
At the early stage of time evolution, $\langle X \rangle$ almost coincides with 
the center of the Gaussian given as the initial condition, and therefore its 
precise calculation has little significance. Accordingly, we do not compute 
$\langle X \rangle$ in the early stage, and as a result, the $\langle X \rangle$ 
shown in Fig.\ref{fig:xvev-kappa-variable} is displayed with the initial portion slightly omitted.
On the other hand, the late-time behavior of $\langle X \rangle$ is required 
to determine the time at which $\langle X \rangle$ reaches the singularity. 
To capture this behavior, we have extended the computational domain and 
performed calculations accordingly.

\section{Time for the singularity formation}
\label{singularity-time}

%%%%%%%%%%%%%%%%%%%%%%%%%%%%%%%%%%%%%%%%%%
\begin{figure}[t]
  \centering
  \begin{subfigure}[t]{0.48\linewidth}
    \centering
    \includegraphics[width=\linewidth]{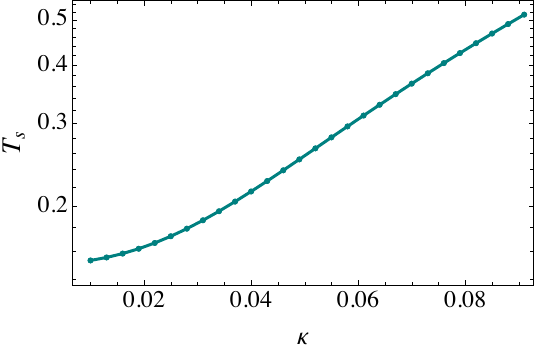}
    \subcaption{Schwarzschild}
  \end{subfigure}\hfill
  \begin{subfigure}[t]{0.48\linewidth}
    \centering
    \includegraphics[width=\linewidth]{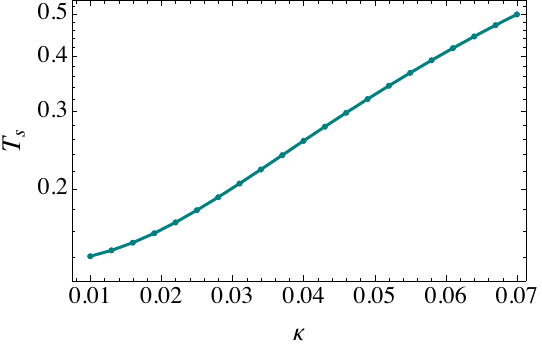}
    \subcaption{Schwarzschild-dS}
    \vspace{0.8em}
  \end{subfigure}
  \makebox[\linewidth]{%
    \begin{subfigure}[t]{0.48\linewidth}
      \centering
      \includegraphics[width=\linewidth]{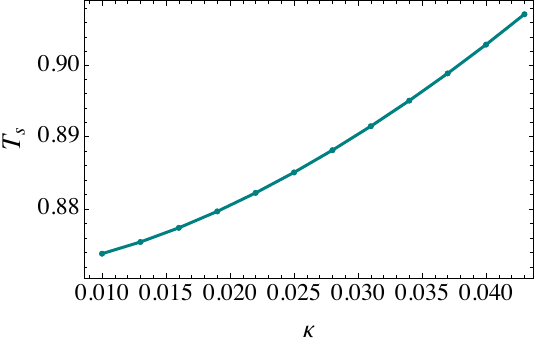}
      \subcaption{Planar-AdS}
    \end{subfigure}
  }
  \caption{The time $T_s$ for the singularity formation as a function of $\kappa$ for the Schwarzschild, and Schwarzschild-dS black hole, and topological AdS black holes with planar horizon.}
  \label{fig:time-singularity-appendix}
\end{figure}
%%%%%%%%%%%%%%%%%%%%%%%%%%%%%%%%%%%%%%%%%

In Fig.~\ref{fig:time-singularity-appendix}, we present the behavior of $T_s$ as a function of $\kappa$ for Schwarzschild, Schwarzschild-de Sitter, and topological AdS black holes with a planar horizon. The same parameter values as those used in Figures~\ref{fig:closed-kappa-variable} and~\ref{fig:flat-open-kappa-variable} are set. The Schwarzschild-AdS black hole case is shown in Fig.~\ref{fig:time-singularity} of the main text, while the results for other cases are now presented.

As discussed in the last paragraph of section~\ref{section3}, the WDW equation can exhibit a tachyonic instability.  This issue becomes particularly significant for the topological AdS black hole, namely for $\Lambda < 0$ and $k = -1$.  In this case, the potential $\mathcal{V}$ becomes negative, $\mathcal{V} < 0$, in the region $|U| < \sqrt{k/\Lambda}$, leading to an exponential growth of the wave function along the time direction.  Therefore, for the topological AdS black hole with a hyperbolic horizon, we restrict our analysis to the domain where such exponential growth does not play a dominant role. Consequently, we have not extended the numerical computation up to the point where the expectation value of $X$ reaches the singularity, and thus the value of $T_s$ has not been displayed.

%%%%%%%%%%%%%%%%%%%%%%%%%%%%%%%%%%
\bibliography{refs}
%%%%%%%%%%%%%%%%%%%%%%%%%%%%%%%%%%

\end{document}